\documentclass[aps,pre]{revtex4}
\usepackage{graphicx}

\begin{document}

\title{About the Yukawa model on a lattice in the quenched approximation }

\author{Feliciano de Soto}
\email[]{fcsotbor@upo.es}
\affiliation{Departamento Sistemas F\'{\i}sicos, Qu\'{\i}micos y Naturales, 
U. Pablo de Olavide, 41013 Sevilla, Spain}

\author{Jean-Christian Angl\`es d'Auriac}
\email[]{dauriac@grenoble.cnrs.fr}
\affiliation{Institut Neel, 25 avenue des Martyrs, BP 166 Grenoble Fr-38042 }
\affiliation{Laboratoire de Physique Subatomique et Cosmologie, 53 avenue des Martyrs,Grenoble Fr-38026 }

\begin{abstract}
The Yukawa model in the quenched approximation is expressed 
as a disordered statistical mechanics model on a 4-dimensional 
Euclidean lattice. We study this model.
A particular attention is given to the singularities
of the Dirac operator in the phase diagram. A careful analysis of
a particular limiting case shows that finite volume effects can be huge and
questions the quenched approximation. This is confirmed by a Monte-Carlo 
simulation in this limiting case and without the quenched approximation.
We include also some results concerning the symmetries of this model.
\end{abstract}

%\pacs{05.50+q}

\maketitle

\section{Introduction and model}

\subsection{Introduction}

Since long it has been recognized that a Quantum Field Theory can
be expressed as a Statistical Mechanics problem. The two areas have
benefited from this proximity and many techniques developed in one
context have then been used in the other~\cite{zinn}. The expression of 
quantum field theories 
as statistical mechanical problems has been especially useful in the
context of the fundamental theory describing the interaction of 
quarks and gluons {\it i.e.} Quantum ChromoDynamics  (QCD), where usual perturbative
techniques fail. Indeed it has been found that perturbative series in any 
quantum field theory have a zero
convergence radius and are asymptotic but never convergent~\cite{zeroradius}. In such
situations, it is common to resort to a numerical approach based on 
the Feynman path integral formulation, where the system is described
by a discretized action on a space-time lattice~\cite{wilson}.

The numerical formulation of QCD on a lattice  is nowadays among 
the most challenging problems of numerical   
physics and the progress have been very important during the last decades. %~\cite{progress_qcd}. 
The methods developed in this context can also be  applied to
other quantum field theories~\cite{rothe} in situations where perturbation theory fails,
for example in the investigation of binding energies. Indeed, such calculations 
would require the evaluation of an infinite number of contributions in 
a perturbative scheme. 

In this paper we study the simplest fermion  quantum field theory in 
four space-time dimensions, that is the model introduced by H. Yukawa for 
the nuclear interaction \cite{yukawa}. 
The model is described in detail in reference \cite{paper1}, but
since the aim of this paper is to adopt a statistical mechanics point
of view, we simply sketch extremely schematically how one goes from
the nuclear physics modelization to the statistical mechanics formulation.

Similar models were analyzed some time ago using the same 
techniques~\cite{Lee}, and two distinct regimes were found: 
for small and large values of the coupling constant the system
was numerically solvable while for intermediate values it was not.
In this paper we shall address in detail this issue, also found 
in~\cite{paper1}. The same techniques have been used for a numerical 
study of a similar model~\cite{Gerhold}, where
some bound on the Higgs boson mass is established based on a Yukawa 
coupling between quarks and the Higgs boson.

\subsection{The model}

The model introduced by Yukawa aimed at a description of nuclei via 
the exchange of massive particles
in analogy with Quantum Electrodynamics, except that the 
particles mediating the nuclear force
have to be massive in order to have a finite range interaction. 
Although we assume QCD is the fundamental
theory of quark interactions responsible for nuclear interactions,
one boson exchange models are still mandatory in the 
nuclear physics community.

Therefore, in the Yukawa model, nucleons and mesons are considered 
elementary particles 
--i.e. without an internal structure--, 
represented by local fields. The mesons are bosons 
represented by a complex scalar field $\phi$ while nucleons 
are fermions represented by 
a four component grassmannian Dirac spinor $\psi$.
To get a statistical mechanics model one works in an 
Euclidean space instead of a Minkowski space, this is achieved 
by performing a  Wick 
rotation~\cite{montvay} and the space-time is discretized into a 
four-dimensional
hyper-cubic lattice. %Therefore if $N$ is the number of sites of the lattice
%the action is represented by a $4N \times 4N$ matrix. 
One possible choice for the discretized action~\cite{montvay} is
\begin{equation}\label{action}
S = \frac{1}{2} \sum_x \left[\left(8+\mu^2\right)\phi_x^2 -
 2\sum_\nu\phi_x\phi_{x+\nu}\right] + 
 \sum_x \bar\psi_x D_{\rm w} \psi_x +
 g\sum_x \bar\psi_x\phi_x\psi_x
\end{equation}
which is the sum of three terms $S = S_{\rm KG} + S_{\rm W} + S_{\rm I}$.
In the first term, which is just a Klein-Gordon action for a free bosonic field,
$x$ runs over the $N$ sites, $\nu$ runs over the four space-time direction,
and $\mu$ is the meson mass.
The second term is a bilinear in the Dirac-Wilson operator $D_{\rm w}$. 
It is a $4N \times 4N$ matrix,
with elements 
\begin{equation}
\left(D_{\rm w}\right)_{x\, y}= 1_{4} \delta_{x,  y}  -\kappa \sum_{\nu}\left(
(1_{4} + \gamma_\nu) \delta_{x,  y-\nu} +
(1_{4} - \gamma_\nu) \delta_{x , y+\nu} 
\right)
\end{equation}
where $1_{4}$ is a $4\times4$ unity matrix and $\gamma_{\nu}$ are the
Dirac matrices ($\bar\psi$ is the conjugate of $\psi$),
$\kappa$ is the so-called hopping parameter related to the bare 
fermion mass $M=1/\kappa-8$.
The coupling between the two fields is realized in the simplest way by the third
term where $g$ is the coupling constant.
Every  dimensional quantity has been redefined in terms of the 
lattice spacing $a$, therefore the model depends on the three adimensionalized 
lattice parameters $g$, $\mu$ and $\kappa$. It depends also on the size 
of the lattice. 
In this work we use periodic boundary conditions and take the four 
dimension equal.

Propagators in Quantum Field Theory are expressed using Wick contractions. 
From the statistical mechanics 
point of view it amounts to computing expectation values and to combining 
them together.  
For example the elementary fermion propagator reads:
\begin{equation}
S(x,y)  = \frac{1}{Z}\int\left[d\phi\right]
\left(D(\phi)^{-1}\right)_{xy}\det\left(D(\phi)\right)e^{-S_{\rm KG}(\phi)}
\label{eq:corr}
\end{equation}
where $x$ and $y$ are two sites of the lattice and 
\begin{equation}
\label{eq:Dirac}
D(\phi) = D_{\rm w} + g\phi   
\end{equation}
is the interacting Dirac operator.
$Z$ is the normalization factor of the probability
distribution of the fields and it is not calculated in practice.
Propagators like (\ref{eq:corr}) provide a simple way of computing 
the renormalized mass $m$ of an interacting particle in a QFT, as 
\begin{equation}
 C(x_4) = \sum_{x_1,x_2,x_3} S(x,0) \sim \cosh{m \left(\frac{L}{2}-x_4\right)}
\end{equation}
where $x_4$ is the time coordinate.  The calculation of 
renormalized masses is performed by 
producing the fields $\phi_x$ according to a joint probability distribution:
\begin{equation}
\Pi\left(\left\{ \phi_{x}\right\} \right)\sim\det\left(D(\phi)
\right)e^{-S_{\rm KG}(\phi)}\label{eq:proba}\ 
\end{equation}
and computing $S(x,0)$ as the average over field configurations of $\left(D(\phi)^{-1}\right)_{0y}$. Note that it implies solving a linear system, not
a full inversion of the Dirac operator.
\medskip

In the bosons probability distribution Eq.~\ref{eq:proba} the evaluation of
the fermionic determinant is --by large-- the most expensive part
of the calculation. Sophisticated methods have been developed for dealing with 
this difficulty, as Hybrid Monte Carlo simulations~\cite{hmc}, 
but the study of the
model neglecting the effect of the determinant on the weight of field 
configuration, called quenched approximation, deserves interest yet
and will be described in some detail in the next section.
Finally, let us remind that to extract physical quantities
one needs to be as close as possible to a critical point 
so that the operator $D(\phi)$ has low modes.
This implies numerical difficulties as in the vicinity of any critical point.

\section{The quenched approximation}

The quenched approximation consists in neglecting the variation 
of $\det\left(D(\phi)\right)$ among the field configurations.
From a physical point of view, this determinant accounts for the
creation of virtual nucleon-anti-nucleon pairs, and its effect is
expected to be small as long as meson mass is smaller than nucleon
one. It simplifies considerably the problem since
now Eq. \ref{eq:proba} becomes
\begin{equation}
\mbox{Prob}\left(\left\{ \phi_{x}\right\} \right) \sim 
\exp\left(-\frac{1}{2}\sum_{x}\left[\left(8+\mu^{2}\right)\phi_{x}^{2} - 
2\sum_{\mu}\phi_{x+\mu}\phi_{x}\right]\right)\label{eq:probaphi}
\end{equation}
This distribution does not anymore involve the Dirac operator and
is easy to implement. Indeed the quadratic form in the exponential
can be diagonalized straightforwardly, simply going to the discrete
Fourier space. We note $\tilde{\phi}_{k}$ the Fourier transform of
the $\phi_{x}$. The $\tilde{\phi}_{k}$ are complex and their
joint probability factorizes
\begin{equation}
\mbox{Prob}\left(\left\{ \tilde{\phi}_{k}\right\} \right) \sim 
\prod_{k}\exp\left(-\frac{1}{2}
\frac{\left|\tilde{\phi}_{k}\right|^{2}}{\sigma_{k}^{2}}\right)
\end{equation}
with $\sigma_{k}^{2}=\frac{1}{\mu^{2}+\sum_\nu\hat{k_\nu}^{2}}$ where 
$\hat{k_\nu}=2\sin\frac{k_{\nu}}{2}$
with the extra constraint $\tilde{\phi}_{k}^{\star}=\tilde{\phi}_{-k}$
in order to get real values for $\phi_x$. It is then simple to draw 
independently the real and imaginary part
of each $\tilde{\phi}_{k}$ (for $k>0$) from a centered Gaussian
distribution with variance $\sigma_{k}^{2}$. The partial
distribution of the $\phi_{x}$ ({\it i.e.} integrating out all $\phi_{y}$
but $\phi_{x}$) is also a Gaussian with a variance $\sigma$ independent
of $x$ and given by 
\begin{equation}
\sigma^{2}(\mu)=\frac{1}{N} \sum_{k}\sigma_{k}^{2} =
\frac{1}{N}  \sum_{k}\frac{1}{\mu^{2}+\sum_\nu\hat{k_\nu}^{2}}
\label{eq:variance}
\end{equation}
In summary the $\phi_{x}$ are Gaussian dependent with the same variance
and the $\tilde{\phi}_{k}$ are independent with a variance depending
on $k$. 

It is straightforward to compute analytically the meson correlator 
\begin{eqnarray*}
C(t) & = & \left\langle \sum_{x,y,z}\phi(x_{1},x_{2},x_{3},x_{4})\phi(x_{1}+x,x_{2}+y,x_{3}+z,x_{4}+t)\right\rangle \\
 & = & \frac{1}{L}\sum_{k_{4}}e^{ik_{4}t}\frac{1}{\mu^{2}+\hat{k}_{4}^2}
\sim \cosh \mu\left(\frac{L}{2}-t\right)
\end{eqnarray*}
($C$ does not depend upon the $x_\nu$'s due to translational invariance)
however when the bosons do interact, directly or through
the fermions when quenched approximation is not assumed, the analytical
calculation is not possible and one has to perform numerical calculation
sampling the field configurations in order to get the re-normalized meson mass. 
In this context, three estimators of the correlator $C(t)$ are possible. 
For the first estimator $C_{0}(t)$ the point
$(x_{0},y_{0},z_{0,},t_{0})$ is fixed and can be chosen to be the origin, for the
second estimator $C_{1}(t)$ the point $(x_{0},y_{0},z_{0,},t_{0})$
runs over all the points of the time-slice $t_0=0$, and for the third
one $C_{2}(t)$ all pairs of lattice points are considered. Obviously
these three estimators give the same average as it should due 
to the translational 
invariance, but their variances are very different. 
In appendix \ref{appendixB} we 
give the three expressions of the variance corresponding to 
the three estimators. 
We see that only $C_{2}(t)$ is self-averaging. For the first
estimator the variance diverges with the size of the lattice, while
for the second it goes to a finite value. It means that \emph{only}
with the third estimator larger sizes imply less configurations in
the average. Consequently for the case of interacting bosons, in
the unquenched calculation for example, the third estimator $C_2$
should be considered.

\section{Symmetries }

In this section we discuss some symmetries of the Dirac 
operator with Yukawa coupling Eq.\ref{eq:Dirac}. Being
associated to the action, these
symmetries hold in both quenched and unquenched calculations. 
They are interesting {\it per se} but also useful for 
numerical treatment.

\subsection{Symmetries holding separately on each boson configuration}
Let us first note that, using the representation for the Dirac 
matrices~\cite{montvay}, the operator 
$J=\imath \gamma_1 \gamma_3$ is an involution verifying
$J\gamma_{\nu}J=\mbox{transpose}(\gamma_{\nu})$ for the four
Dirac matrices $\gamma_{\nu}$. It is
then straightforward to verify that 
\begin{equation}
D=JD^{\star}J\label{eq:DJD}
\end{equation}
where $D^{\star}$ is the complex conjugate of $D$. Let $V$ be an
eigenvector of $D$ belonging to the eigenvalue $\lambda$. Introducing
the complex conjugation operator $K$, the vector $W=JK(V)$ is also
an eigenvector of $D$ belonging to the eigenvalue $\lambda^{\star}$.
Indeed $DJK(V)=JD^{\star}K(V)=JK(DV)=JK(\lambda V)=\lambda^{\star}JK(V)$.
Moreover $V$ and $W$ are orthogonal. So the eigenvalues appear in
pair of conjugate values and therefore the determinant is never negative.
This non-negativity property is useful to perform hybrid Monte-Carlo
simulation in the unquenched calculation.

The relation Eq \ref{eq:DJD} has another useful consequence. In
Eq. \ref{eq:corr} $S(x,y)$ is a $4\times4$ matrix where row and
column are indexed by the spin at sites $x$ and $y$. To
compute $S(x,y)$ one solves for the propagator $X_{\sigma}$ the
four linear system with the four right hand-side (source term) $Y_{\sigma}$
\begin{equation}
\label{sys}
DX_{\sigma}=Y_{\sigma}
\end{equation}
corresponding to the 4 spin states $\sigma=0,1,2,3$. 
The $4\times4$ matrix $S(x,y)$ is obtained selecting the proper elements 
of the four vectors $X_\sigma$.
We will show
a relation between $X_{1}$ and $X_{2}$, obviously this relation
hods also between $X_{3}$ and $X_{4}$. Indeed it is readily verified,
using sing Eq. \ref{eq:DJD} , that $D\left(\imath JK(X_{1})\right)=Y_{2}$,
in other words 
\begin{equation}
X_{1}=-\imath JK(X_{2})
\end{equation}
and consequently each correlation matrix, and for any field configuration,
has the following form
\begin{equation}
S=\left(\begin{array}{cccc}
a & b & c & d\\
-b^{\star} & a^{\star} & -d^{\star} & c^{\star}\\
e & f & g & h\\
-f^{\star} & e^{\star} & -h^{\star} & g^{\star}
\end{array}\right)
\end{equation}
and the trace of any of these matrix is simply $2\left(\mathcal{R}(a)+\mathcal{R}(g)\right)$.
Note that this form of the correlation matrix holds also for any composite
particle correlator (even using the so-called smeared source).

\subsection{Symmetries holding on the average}
We now show that another simplification appears when averaging the
correlation matrices over the fields configurations. Let us introduce
the automorphy group of the lattice, {\it i.e.} permutations $\pi$ of the
sites of the lattice such that the images of two neighboring sites are also
two neighboring sites. For any such permutation the two fields configurations
$\phi_{x}$ and $\phi_{\pi (x)}$ have the \emph{same} probability,
since both the fermions and the bosons actions are invariant under
the permutation $\pi$. Note that this equality also
holds without the quenched approximation. Let's denote $\pi_{1}$
the particular permutation defined by
\begin{equation}
\pi_{1}(x_1,x_2,x_3,x_4)=(-x_1,x_2,x_3,x_4)
\end{equation}
it is clear that $\pi_{1}$ belongs to the automorphy group. We also
introduce $\pi_{2}$, $\pi_{3}$ and $\pi_{4}$ corresponding respectively
to $x_2,$, $x_3$ and $x_4$. 
We have
\begin{equation}
S(x,\phi_{x})=\gamma_{5}\gamma_{k}S\left(\pi_{k}(x),\phi_{\pi_{k}(x)}\right)\gamma_{k}\gamma_{5}
\end{equation}
where $x=(x_1,x_2,x_3,x_4)$ and $k=1,2,3,4$, $\gamma_{5}=\gamma_{1}\gamma_{2}\gamma_{3}\gamma_{4}$
(see ref \cite{zinn} section 8.2 for the free fermion case, the extension to
the Yukawa model treated in this paper is straightforward). 
Using this relation the 1-fermion correlation matrix,
when the source is located at the origin, takes the forms
\begin{equation}
\label{matS}
C(x_4)= \sum_{x_1,x_2,x_3} S(x_4) = \left(\begin{array}{cccc}
c(x_4) & 0 & 0 & 0\\
0 & c(x_4) & 0 & 0\\
0 & 0 & c(L_4-x_4) & 0\\
0 & 0 & 0 & c(L_4-x_4)
\end{array}\right)
\end{equation}
the precise form~\ref{matS} obviously 
depends on the chosen representation for the Dirac matrices,
but in any representation the matrix $C(t)$ depends on a single 
function $c(t)$ instead of 16 functions.

\section{The Dirac operator spectrum in the phase space $\kappa-g$}

Let us recall that the model depends on three independent parameters,
$\kappa$, $g$ and $\mu$.  As shown above, in the \emph{quenched approximation} 
the probability of a $\phi_{x}$ depends only on $\mu$ and not on $\kappa$ or $g$, 
it is the same everywhere in the parameter space. 
In this section we work at constant value of $\mu\sim0.1$

Any numerical computation of a physical
quantity will imply some inversions of the Dirac operator Eq. \ref{eq:Dirac}.
We know that this inversion will have to be performed with values
of $g$ and $\kappa$ such that the linear system is difficult to
invert. In practice, in some region of the $g-\kappa$ plane and
for a given value of the linear sizes of the lattice, solving for $X$ the system
$DX=Y$ will not be possible. Indeed, depending on the numerical method
used, either the algorithm will not converge, or it will find a wrong
solution. To quantify how ill conditioned the linear system is, it
is customary to use the condition number. By definition a condition
number measure how the solution of the system changes when the RHS
term changes~\cite{saad}. With the appropriate choice of the norms
the condition number is the ratio 
$r=\frac{|\lambda_{a}|}{|\lambda_{i}|}$
of the largest to the smallest modulus of the eigenvalues. With this
definition, and for the type of system we consider, a system can be
inverted reasonably if the condition number is smaller than $100\sim1000$.
Note however that a condition number can be arbitrarily large but still
the system is invertible. This is the case if the RHS of the system
is in the kernel of the operator. This situation occurs with some
preconditioning. 

We now note that, due to the specific form of the Dirac operator Eq.
\ref{eq:Dirac}, one has 
\begin{equation}
D(\alpha g,\alpha\kappa;\phi_{x})-1=\alpha\left(D(g,\kappa;\phi_{x})-1\right)
\end{equation}
where $1$ denotes the $4N\times4N$ unity matrix. Since the probability
of the $\phi_{x}$'s does not depend on $g$ and $\kappa$ one is
lead to introduce the polar coordinates $r$ and $\theta$ of the
parameter space ($g=r\cos(\theta)\;\kappa=r\sin(\theta)$. For a given
value of $\theta$ the spectrum of $D$ evolves straightforwardly
: the eigenvectors are then left unchanged and the eigenvalues $\lambda_{k}$
evolve according to
\begin{equation}
\lambda^{k}(r,\theta)=\frac{r}{r_{0}}\lambda^{k}(r_{0},\theta)+1-\frac{r}{r_{0}}\label{eq:prince}
\end{equation}
In a spectral decomposition of $D$, varying $r$ only changes the
relative weights of the eigensubspaces. The value of $\theta$ fixes
the spectrum, and the value of $r$ the relevant part of the spectrum.
In general the eigenvalues are complex $\lambda^k=\lambda^k_{\rm{R}}+\imath\lambda^k_{\rm{I}}$.
Let us give a fixed value to $\theta$ and denote the spectrum $\lambda^k(r)$.
We choose a reference value $r_0$ (one can take for example $r_0=1$) and 
note $\Lambda^k = \lambda^k(r_0)$, one has
\begin{equation}
\left|\lambda^{k}(r)\right|^{2}=
\left(\left|\Lambda^{k}\right|^{2}-2\Lambda_{\rm{R}}^{k}+1\right)r^{2}
+2\left(\Lambda_{\rm{R}}^{k}-1\right)r+1
\label{eq:lambdader}
\end{equation}
So the modulus of the each eigenvalue is a parabola as a function
of $r$. All these parabola intersect at the point ($r=0,\lambda=1)$.
They also intersect each other at others points, and the two extremal
eigenvalues change when $r$ changes(see Fig.~\ref{fig:schema}). 
The eigenvalue labeled by $k$
will reach its smallest value 
\begin{equation}
\label{minev}
m_k = \frac{(\Lambda^k_{\rm I})^2}{|\Lambda^k|^2-2\Lambda^k_{\rm R}+1}
\end{equation}
for $r=\frac{\Lambda^k_{\rm R}-1}{|\Lambda^k|^2-2\Lambda^k_{\rm R}+1}$.
Therefore only the eigenvalues with $\Lambda^k_{\rm R}<1$ and
$\Lambda^k_{\rm I}\ll 1$ give rise to a small denominator in the 
condition number. 
When $r \simeq 0$ the eigenvalue of lowest (resp. largest) modulus will be 
the one with the smallest (resp. largest) value of $\Lambda_{\rm R}-1$,
therefore the condition number increases continuously from the value 1.
In the other limit $r \gg 1$ the eigenvalue of lowest (resp. largest) 
modulus will be 
the one with the smallest (resp. largest) value of
$|\Lambda^{k}|^{2}-2\Lambda_{\rm{R}}^{k}+1$, and the condition
number tends to a finite value (the ratio of the two values above).
In the intermediate regime, the condition number has a very complicated
behavior with a lot of maxima and minima. We analyze this behavior
is the next subsection for different case.

\begin{center}
\begin{figure}
\noindent \centering{}
\includegraphics{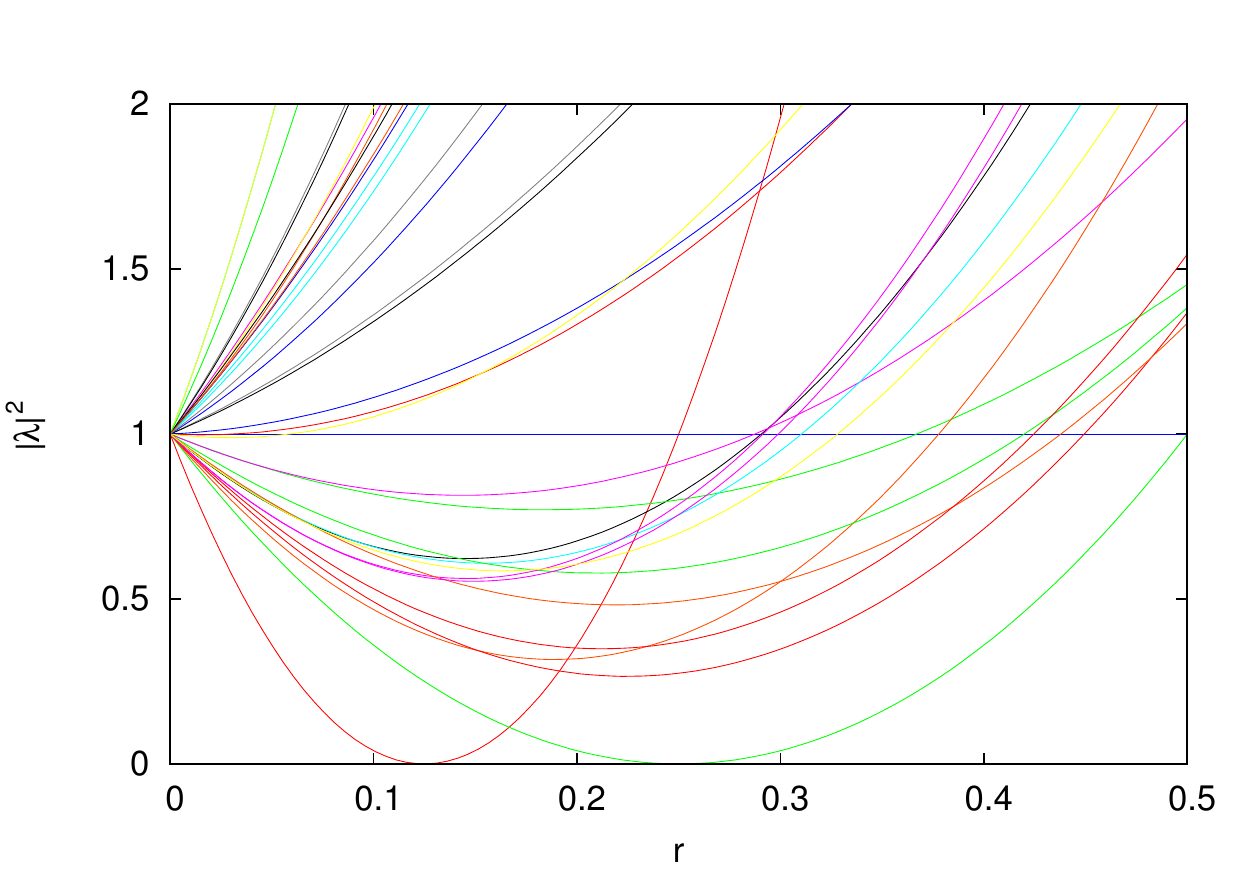}
\caption{Evolution of the square of the modulus 
of some eigenvalues as a function of $r$
for the free case $g=0$ {\it i.e.} $\theta=\frac{\pi}{2}$
\label{fig:schema} }
\end{figure}
\par\end{center}

\subsection{case $\theta=\frac{\pi}{2}$}

This correspond to $g=0$ and therefore this is the trivial case
of \emph{non interacting} fermions, it is included for illustrating purpose. 
The evolution of the spectrum of $D$ as a function
of $r=\kappa$ is straightforward. Performing a discrete Fourier transform
one finds that the eigenvalues are given by
\begin{equation}
\lambda_{k}=
\left(1-2\kappa\sum_{\nu}c_{\nu}\right)\pm2\kappa\imath\sqrt{\sum_{\nu}s_{\nu}^{2}}
\label{eq:eigen_libre}
\end{equation}
with $c_\nu=\cos(k_\nu)$ and $s_\nu=\sin(k_\nu)$. Therefore the 
condition number behaves as
\begin{eqnarray*}
c(r=\kappa)=\left\{ 
\begin{array}{ll}
\frac{1+8\kappa}{1-8\kappa} & \kappa<\frac{1}{8}\\
\frac{1+8\kappa}{8\kappa-1} & \frac{1}{8}<\kappa<\frac{1}{6}\\
\frac{1+8\kappa}{1-4\kappa} & \frac{1}{6}<\kappa<\frac{1}{4}\\
\frac{1+8\kappa}{4\kappa-1} & \frac{1}{4}<\kappa<\frac{1}{2}\\
1+8\kappa & \frac{1}{2}<\kappa
\end{array}\right.
\end{eqnarray*}
The condition number diverges at the two values $\kappa=\frac{1}{8}$
corresponding to $k=(0,0,0,0)$ and $\kappa=\frac{1}{4}$ 
corresponding to $k=(\frac{L_1}{2},0,0,0)$
(or permutation when the corresponding $L_\nu$ are even).
The first value $\frac{1}{8}$ is the critical
value, while the other is unphysical since it corresponds to a negative
mass.

This is compatible with the framework introduced the introduction of
this section. For example Eq~\ref{minev} becomes 
$m_k=\frac{\sum{s_\nu^2}}{(\sum{c_\nu})2+\sum{s_\nu^2}}$.
On figure Fig~\ref{fig:schema} the evolution of some eigenvalues
with $r=\kappa$ is presented. One sees that for zero eigenvalues
appear only for $r=\frac{1}{8}$ and $r=\frac{1}{4}$.

\subsection{\label{sub:casekappa0quenched}case $\theta=0$}

This case corresponds to $\kappa=0$ and describes infinitely heavy fermions. 
It is unphysical but non trivial. 
However it is instructive to study it from a statistical mechanics 
point of view,
and also because if some continuity is to apply, it should not be
very different from the $\theta$ small case. In that case the Dirac
operator is simply diagonal and the $4\times4$ blocs are given by
\begin{equation}
\left(D(\phi)\right)_{x\, y}=\left(1+g\phi_{x}\right)\delta_{xy}1_{4}
\end{equation}
Obviously the eigenvalues $1+g\phi_{x}$ are all degenerated four
times and \emph{real}.
 The Eq. \ref{eq:lambdader} becomes simply
$(\lambda^{k}(r))^2=(1+r(\Lambda^{k}-1))^2$.
Since $\lambda^{k}(r)$ is linear with $r$, 
for any eigenvalue $\Lambda^k$<1 there will be a value of $r^k$ for which
$\lambda(r^k)=0$. This is the worst situation since for any fields
configuration the determinant of the Dirac operator will exactly
vanish $\frac{N}{2}$ times. This is illustrated
on figure~\ref{fig:schema}, where all the eigenvalues as a function
of $g$ are shown. The two eigenvalues of largest 
and lowest modulus are emphasized. One clearly sees
that the eigenvalue of lowest modulus vanishes for many
values of $r$ (recall $r=g$ when $\kappa=0$). This situation 
is in contrast with the previous case $\theta=\frac{\pi}{2}$ 
where the eigenvalue of lowest modulus vanishes only twice
(for $r=\frac{1}{8}$ and $r=\frac{1}{4}$). Here the
eigenvalues are ``protected'' by their imaginary part.

In other words, the $\phi_{x}$ are $N$ correlated real random
variables following the probability distribution Eq. \ref{eq:probaphi},
and we are evaluating the condition number $c(r)$ which is in that
case 
\begin{equation}
c(r;\phi)=\frac{\max\left|1+r\phi_{x}\right|}{\min\left|1+r\phi_{x}\right|}
\end{equation}
 Let us suppose that the $\phi_x$ have been sorted in ascending
order. We note $\lambda_{m}$ the largest negative eigenvalue. Since
$N$ is very large, we assume $m\sim\frac{N}{2}$ and there is at
least one negative and one positive eigenvalue. The schema on figure
Fig \ref{fig:schema_kappa_0} illustrates the behavior of the spectrum of
the Dirac operator for a given $\phi_{x}$ realization. Each eigenvalue
varies linearly with $g$. 
%The modulus of the eigenvalue of largest
%modulus is the same whatever $g$ and it corresponds to the maximum
%of the $\phi_{x}$. 
Therefore the condition number is controlled by
the eigenvalue of smallest modulus, which is a piecewise linear function
of $g$. The selected eigenvalue changes each time
$g$ reaches a value $g_i=-\frac{2}{\phi_{i}+\phi_{i+1}}$,
and reaches zero for $g_{i}^{\star}=-\frac{1}{\phi_{i}}$ for $0\le i\le m$,
which $g^\star_0<g_0<g^\star_1 < g_1 \cdots$. Consequently
three regimes occur. Firstly when $g<g_{0}^{\star}$
the condition number is a continuous increasing function of $g$ (homographic)
which diverges at $g_0$. Secondly in the intermediate
regime $g_{0}^{\star}<g<g_{m}^{\star}$
the condition number varies extremely fast diverging $m$ times.
Finally for $g_{m}^{\star}<g$
the condition number decreases homographically saturating at a finite value.
%$\frac{\phi_{N}}{\phi_{m}}$. 
This is illustrated on Fig.~\ref{fig:schema_kappa_0} where the 
extreme values $g^\star_0$ and $g^\star_m$ are indicated. 

In order to perform analytical evaluation of those tree regimes
we simplify the problem by choosing the fields $\phi_x$ {\it independent}
with zero mean and a variance given by Eq.~\ref{eq:variance}.
It turns out that this simplification does not change substantially
the average value of the eigenvalue of lowest modulus, as it is illustrated
on Fig.~\ref{lambda_min_vs_g_various_L}. This figure shows, among other
things detailed below, the two curves of the eigenvalues of lowest modulus 
(curves labeled $N=131072$) as a function of $r$ 
when the $\phi_x$ are independent identically distributed Gaussian variables
and when they are dependent : the two curves are completely 
indistinguishable.
Within this assumption, when the number $N$ of lattice sites increases 
$\phi_{m}$ goes to zero as
\begin{equation}
\left\langle \phi_{m}\right\rangle   =  -\sigma\int_{0}^{\infty}\left(1-\mbox{erf}\frac{x}{\sqrt{2}}\right)^{N}dx\sim -\sigma\sqrt{\frac{2}{\pi}}\frac{1}{N}
\label{eq:lambdamin}
\end{equation}
Therefore $g_m^\star \sim N$ and the third region shrinks when the lattice 
size increases. In other words the decreasing of the condition number 
for large values of the coupling constant $g$  at $\kappa=0$ is a size effect. 
On the other limit for small $g$, the first region $g<g_0^\star$ is delimited by
the smallest field $\phi_0$ whose average is given by
\begin{eqnarray}
\left\langle \phi_{0}\right\rangle  & = & -\sigma\int_{0}^{\infty}
\left[1-\left(\mbox{erf}\frac{x}{\sqrt{2}}\right)^{N}\right]dx
\end{eqnarray}
we see that $\left\langle \phi_0 \right\rangle$ diverges extremely 
slowly with $N$. 
To have $\left\langle \phi_{0}\right\rangle $ of
the order of $\xi$, one needs a huge lattice  
of $N\sim \xi\exp\frac{\xi^{2}}{2}$ sites.
Therefore the first
region also disappears in the thermodynamical limit. However
this size effect will never be seen in an actual computation.
Finally we conclude that only the second region survives
the large lattice volume. Let us recall that in this region 
and for any fields configuration there are $m\sim\frac{N}{2}$ values of $g$
for which one eigenvalue of the Dirac operator is exactly zero.

\begin{center}
\begin{figure}
\noindent \centering{}\includegraphics{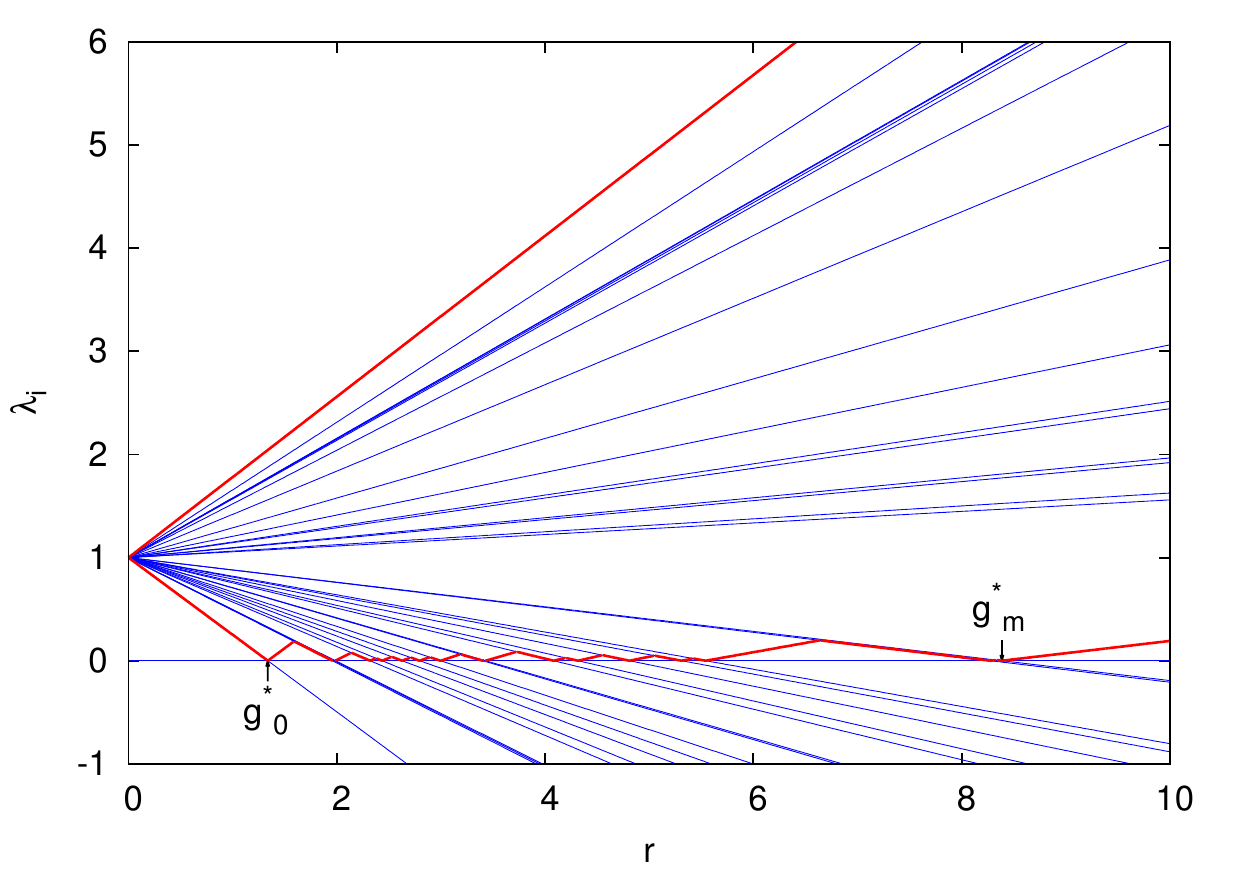}
\caption{Schematic evolution of the modulus of the eigenvalue 
for one $\phi_{x}$ realization for $\kappa=0$. The modulus
of the eigenvalue of largest and smallest modulus are shown in red.
\label{fig:schema_kappa_0} }
\end{figure}
\par\end{center}

In the precedent paragraph the behavior of the condition number for\emph{
a given configuration of the} $\phi_{x}$ has been studied. 
We need now to perform
an average over the realization of the $\phi_{x}$. 
For a fixed value of $g$, different field configurations will give
very different condition numbers, some of them possibly extremely
large. 
Note however that
the condition number is not a physical observable, it is only an indicator
of how difficult the inversion will be. Therefore the most probable value
of the condition number is maybe more sensible. 
From the probability distribution of the $\phi_x$
one can easily compute the average of the smallest 
and largest eigenvalues as a function of $g$. This is done
in Appendix \ref{AppendixA}, the result is
\begin{eqnarray}
\left< | \lambda_i |\right> &\sim&
\frac{1}{N}\sqrt{\frac{\pi}{2}}\sigma\,\exp\frac{1}{2\sigma^{2}} \label{a}\\
\left< | \lambda_a |\right> &\lesssim& g\sigma \sqrt{2\ln N}  \label{aa}
\end{eqnarray}
The eigenvalue of lowest modulus goes to zero 
as $\frac{1}{N}$ but the prefactor increases extremely fast
when $g$ goes to zero. Since $N$ goes to infinity first, for
any non zero $g$, $\left< | \lambda_i |\right>$ goes to zero.
The eigenvalue of lowest modulus increases very slowly with $N$.
This is illustrated in Fig.~\ref{lambda_min_vs_g_various_L}
where we show the eigenvalue of lowest modulus for several lattice volumes.
On this figure and for $N=64\, 32^3=131072$ we have plotted the result
of a ``genuine'' simulation with dependent fields, another simulation
with independent fields, a numerical integration of Eq.~\ref{b},
and the approximation Eq.~\ref{a}. The agreement between these four
calculation is excellent. We have also plotted the eigenvalue of lowest modulus
for other values of $N$ to show the size effects.

The appearance of the three regions described above can be seen
in figure \ref{cn_vs_g}. On this figures we have plotted the
average condition number $\left<\frac{|\lambda_{\rm max}|}{|\lambda_{\rm min}|}\right>$
over 8692 samples as a function of $g$.
If we would have used a smaller discretization of $g$ we would
have even more sharp peaks. The quantity
$\frac{<|\lambda_{\rm max}|>}{<|\lambda_{\rm min}|>}$ is much smoother
since $<|\lambda_{\rm min}|>$ never vanishes, and also
displays the three regimes. Moreover {\em in the quenched
approximation} it makes sense to consider a particular realization
since the weight of a consideration does depend only on $\mu$,
we therefore have plotted a typical configuration.
Finally we have also plotted the average condition number 
without the quenched approximation : this is discussed in the next section.

\begin{center}
\begin{figure}
\begin{centering}
\includegraphics{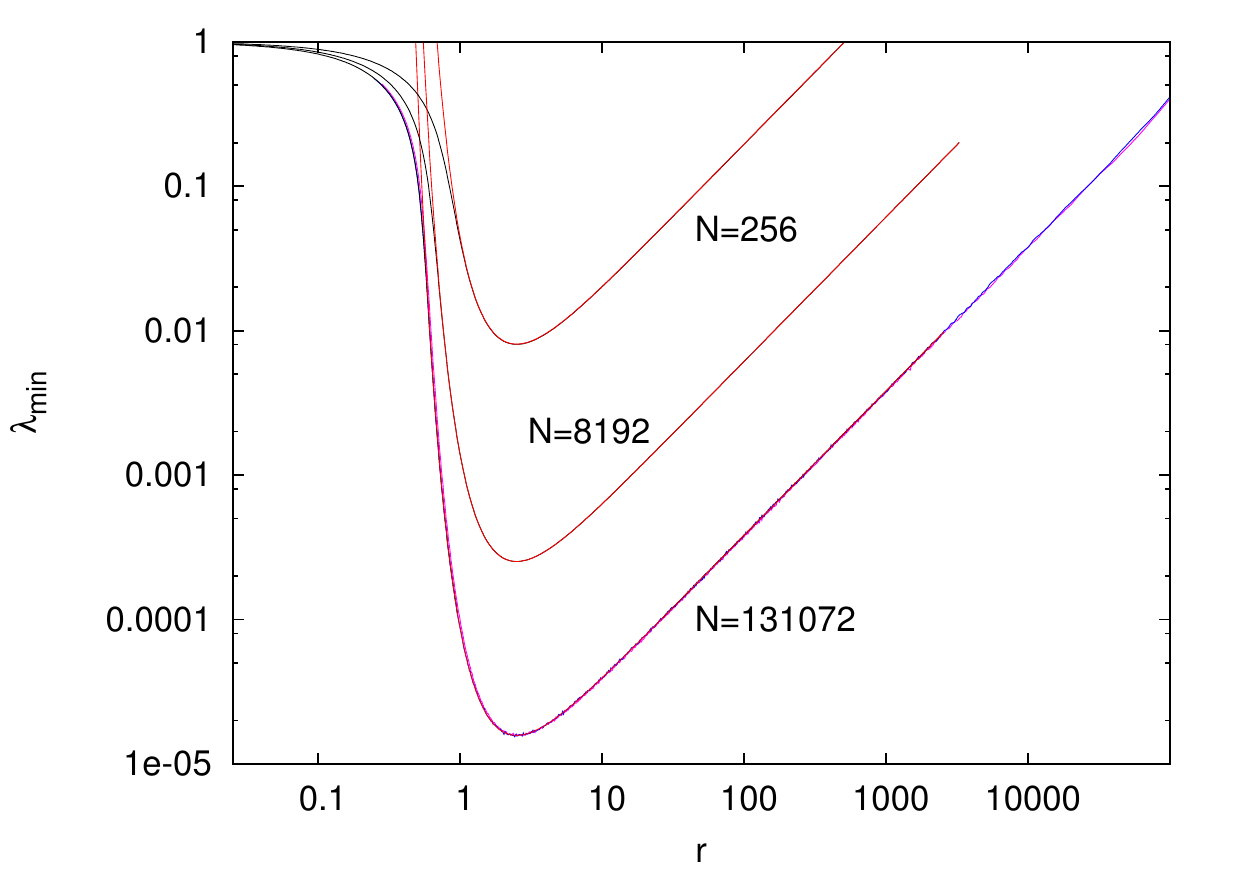}
\par\end{centering}
\caption{Smallest modulus of the eigenvalues {\it vs} $g$ for $\kappa=0$
and three values of $L$. For any value of $L$ the numerical integration
(black) together with the approximation (red) Eq.~\ref{b} are shown.
For the largest size $L=32\times 16^3$ the result of a simulation
averaged over 8692 samples is also shown for both correlated $\phi_x$ 
(magenta) and uncorrelated $\phi_x$ (blue). Actually the curves are
indistinguishable except for small $g<1$ when the approximation of the
integral is not correct.
\label{lambda_min_vs_g_various_L}}
\end{figure}
\par\end{center}

\begin{center}
\begin{figure}
\begin{centering}
\includegraphics{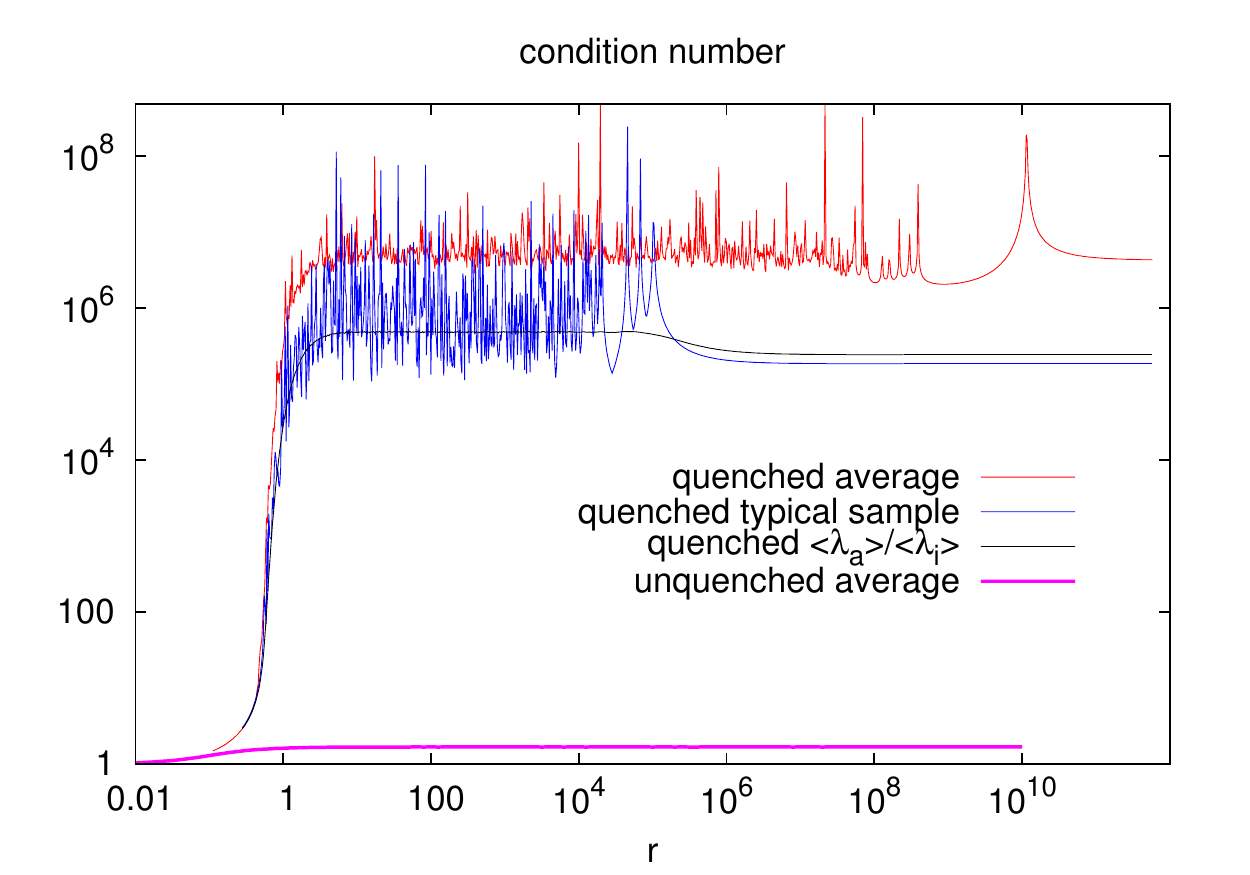}
\par\end{centering}
\caption{Condition number {\it vs} $g$ for $\kappa=0$ for a $32 \times 16^3$
lattice. The quenched average
over 8692 samples is shown in red, and a typical quenched sample is shown
in blue. For comparison the ratio $\frac{|\lambda_a|}{|\lambda_i|}$
is also shown in black. The results of the unquenched Monte-Carlo 
simulation is the thick purple line.
\label{cn_vs_g}}
\end{figure}
\par\end{center}

In conclusion of this subsection, the size effects on this model
for $\kappa=0$ appear extremely severe: 
%, at least with the hypothesis of independent fields. 
for any fixed value of $g$ the
occurrence of configurations with arbitrarily 
small eigenvalues in absolute value
grows with the size. This is reminiscent of the so-called
``exceptional configurations'' which have been encountered
in the context of quenched lattice QCD~\cite{protection}.

\subsection{case $0<\theta<\frac{\pi}{2}$}

This region is non trivial since the Dirac operator cannot
be diagonalized as in the two previous cases. Nevertheless this is
where the physics takes place. As it has been done in Ref.~\cite{paper1},
to perform realistic calculation
one finds the critical line, and one chooses the particular
point close to this line where the ratio of the renormalized masses
of fermion and boson is equal to the physical one. This program
has been done successfully giving consistent results for $g$
small. However for $g$ around 0.7, the linear system Eq.~\ref{sys} 
becomes ill conditioned, preventing any conclusive result.

We have computed the condition number for a typical field
configuration and the result is presented in Fig.~\ref{feli}.
It has been shown in~\cite{paper1} that for small $g$ the critical line, 
defined as the line where the renormalized
mass of the fermion vanishes, is a parabola
originating from the point $\kappa=\frac{1}{8},g=0$. We therefore
expected a diverging condition number along this line. This
is clearly seen on  Fig.~\ref{feli}.
When the coupling $g$ increases it enters in an ill conditioned region where
the condition number shows large fluctuations. The localization
of this ill conditioned region for $\kappa$ small agrees with
what we have shown in Sec.~\ref{sub:casekappa0quenched}. Then
we can ask if this region where the condition number is small
enough is not a size effect, as for the case $\kappa=0$.

We see three possible origins for this problematic region.
It can be that quenched approximation is not working for
those values of $g$. This seems intuitively reasonable since
the determinant in Eq~\ref{eq:proba} precisely give a low weight to
these configuration with a large condition number. Another
possible reason could be the specific choice of the action
and the discretization of the fermion. Finally there is the
possibility that this a fundamental problem of the Yukawa model.  

\begin{center}
\begin{figure}
\centering{}\includegraphics{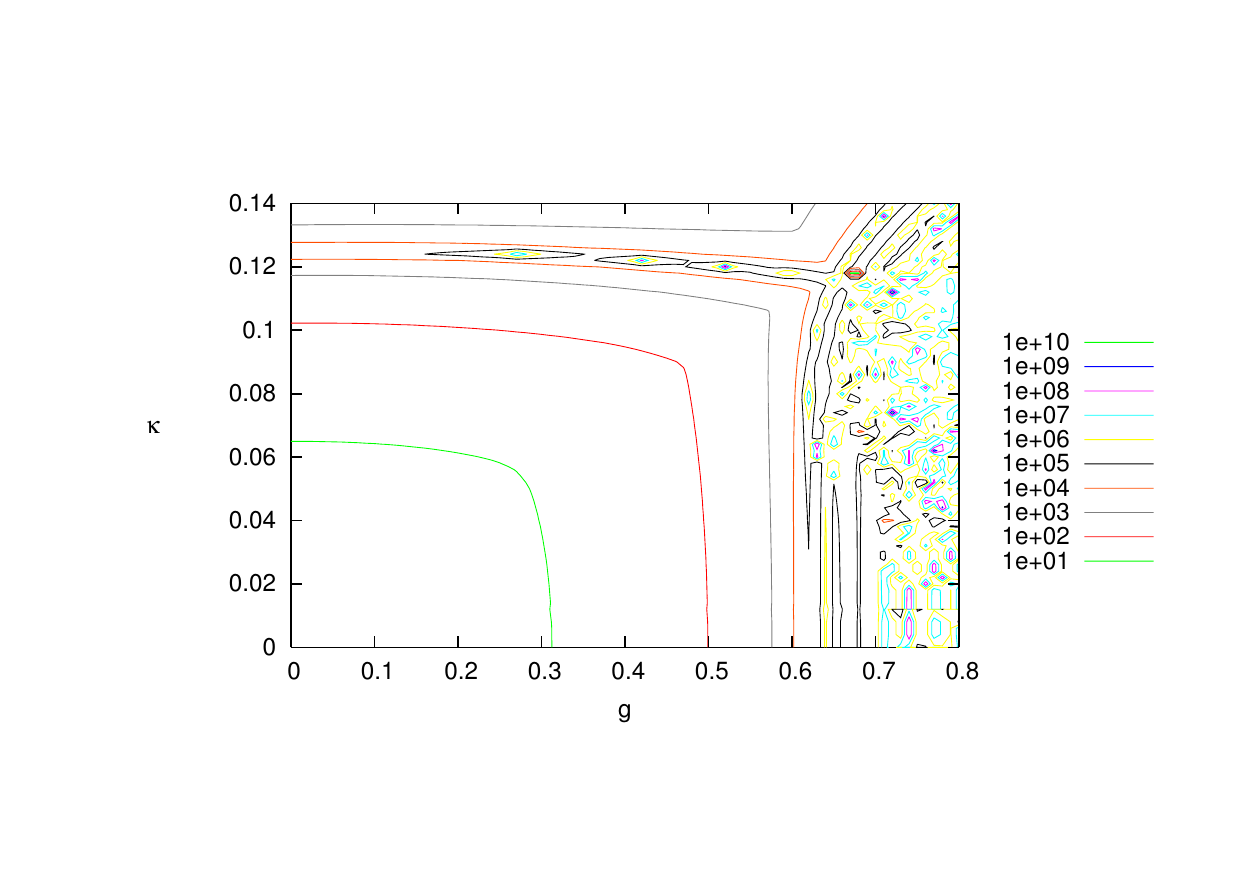}
\caption{Iso-condition number line on a $16^4$ lattice.
\label{feli}}
\end{figure}
\par\end{center}

\section{The $\kappa=0$ case without the quenched approximation}

In this section we consider the simple case $\kappa=0$ as in the
subsection sec. \ref{sub:casekappa0quenched}, but {\em without 
the quenched approximation}. The purpose is to illustrate on this
simple case the consequence of the quenched approximation. 
Intuitively the determinant in the probability density Eq.~\ref{eq:proba}
of the $\phi_x$ gives a vanishing weight to the non invertible configurations.
So we can expect that the configurations to include in the sampling
will not have a large condition number.
But it is possible to have a large determinant, and still
a small eigenvalue, for example if one eigenvalue is small
and all the others are large. These configurations would have a 
non vanishing weight, but still a very large condition number.

The joint probability of the fields $\phi_{x}$ Eq.~\ref{eq:proba} can
be written as
\begin{equation}
\Pi\left(\left\{ \phi_{x}\right\} \right)\sim\exp\left(-\frac{1}{2}\sum_{x}\left[\left(8+\mu^{2}\right)\phi_{x}^{2}-2\sum_{\mu}\phi_{x+\mu}\phi_{x}\right]+4\ln\left|1+g\phi_{x}\right|\right)
\label{joint_prob}
\end{equation}
Since this expression cannot be factorized we have written a simple
Monte-Carlo algorithm to generate the $\phi_x$'s. We use the simple
metropolis algorithm \cite{metropolis}. The normalization factor of 
Eq.~\ref{eq:proba} is very difficult to compute, but the ratio
of the probability of two $\phi$ configurations is very simple to compute
(see Eq.~\ref{joint_prob}). The Monte-Carlo method use this fact to
construct a Markov chain which has the desired distribution
as a fixed point. In practice, we start from a initial
$\phi_x$ configuration, then we choose at random a site $x$
and try to change the value $\phi_x$ for $\phi_x+d\phi$ where $d\phi$
is a random number normally distributed. We accept this change with
the probability $\min(1,\Delta E)$ where $E$ is the variation of 
the argument of the exponential in Eq.~\ref{joint_prob}.
We do not have a proof that this algorithm converge, the difficulty being
that the number of states of the Markov chain is infinite. However 
for all practical purpose it works properly if one choose always
as a starting distribution for a value $g$
an equilibrium distribution for a close smaller value $g-\delta g$.
This indicates that the energy landscape is complicated, probably with
metastable states. This naive algorithm is much simpler than
the well known hybrid Monte-Carlo algorithm \cite{hmc}, but it is
sufficient for our purpose here. We have compare the two algorithms
finding that hybrid Monte-Carlo algorithm is more efficient 
than the naive Monte-Carlo if the parameters are properly chosen,
but they both give the same results with a good accuracy.

Before analyzing the condition number, let us look at the mean
value $\left< \phi_x \right>$ (vacuum expectation value). Let us
first recall that in the quenched approximation, due to 
the symmetry of Eq.~\ref{eq:probaphi} the average value  
$\left< \phi_x \right>$ is zero. This not the case without
the unquenched approximation as seen on Fig.~\ref{phi_moy_unq},
even for small $g$. Indeed performing a $g$-expansion of
Eq.~\ref{joint_prob} one finds that for any site $x$
\begin{equation}
\left< \phi_x \right>_g = g \sum_y \left< \phi_x \phi_y \right>_{g=0} + O(g^3)
= \frac{g}{\mu^2} + O(g^3) 
\end{equation}
The insert of Fig.~\ref{phi_moy_unq} shows the slope  
$\frac{1}{\mu^2}$ at the origin : the agreement is very good. 
Since the average 
%$\left< \phi_0 \right>_g$ grows with $g$, it seems likely that
$\left< \phi_x \right>$ grows with $g$, it seems likely that
$\min(|1+g  \phi |)$ will not easily become small. 
This is indeed confirmed in Fig.~\ref{lambda_min_unq} where we have
plotted the eigenvalues of minimum and maximum modulus 
for both the quenched and unquenched case. It is clearly seen
that $\lambda_{\rm min}$ is never small. Finally the average condition
number is plotted on Fig.~\ref{cn_vs_g} where the drastic
effect of the quenched approximation is clearly seen : a
reduction by six order of magnitude of the condition number.
This reduction is larger with larger lattice. We conclude
that there is no ill conditionned point on this $\kappa=0$
line without the quenched approximation, whereas it is 
everywhere ill conditionned in the quenched approximation.

\begin{center}
\begin{figure}
\centering{}\includegraphics{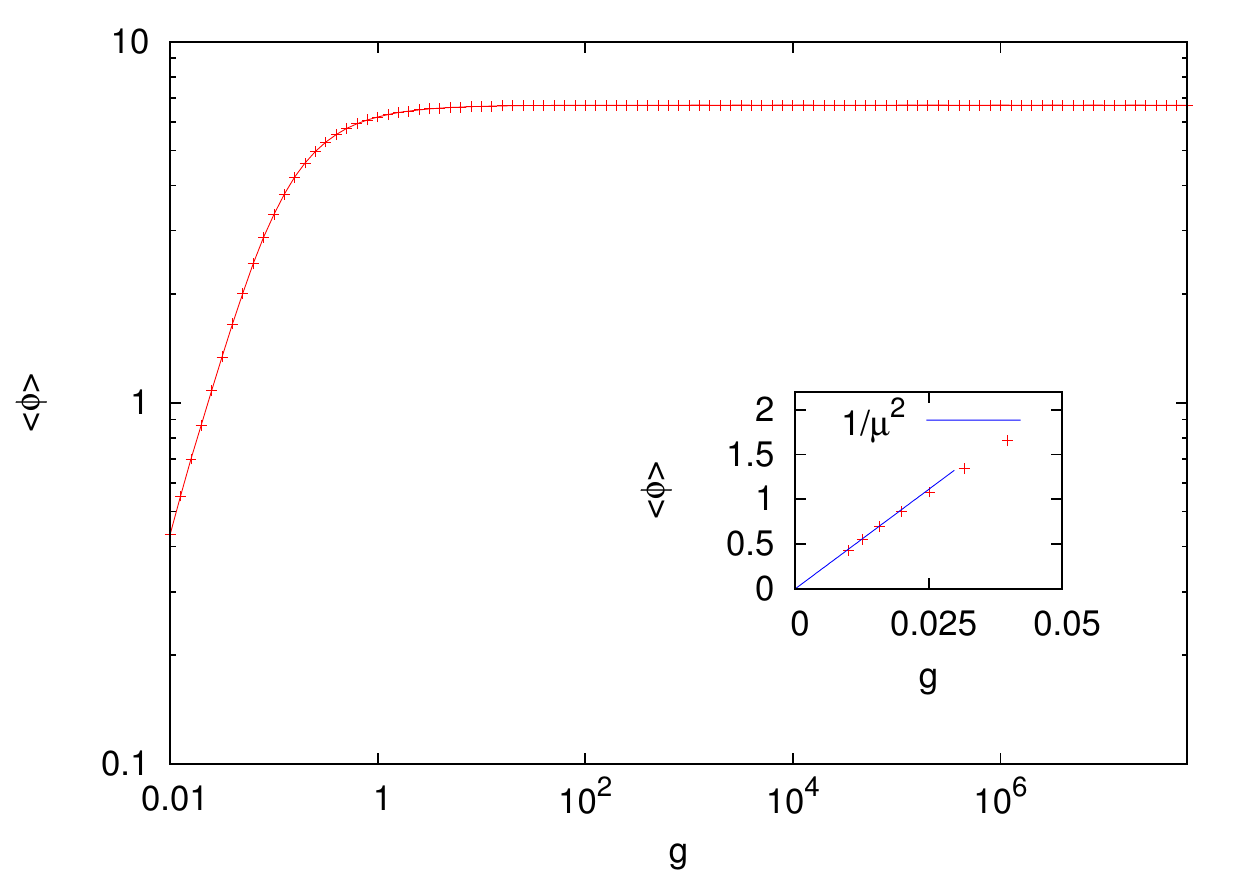}
\caption{$<\phi>$ in an unquenched simulation on a $32 \times 16^3$ lattice
for $\kappa=0$ \label{phi_moy_unq}}
\end{figure}
\par\end{center}

\begin{center}
\begin{figure}
\centering{}\includegraphics{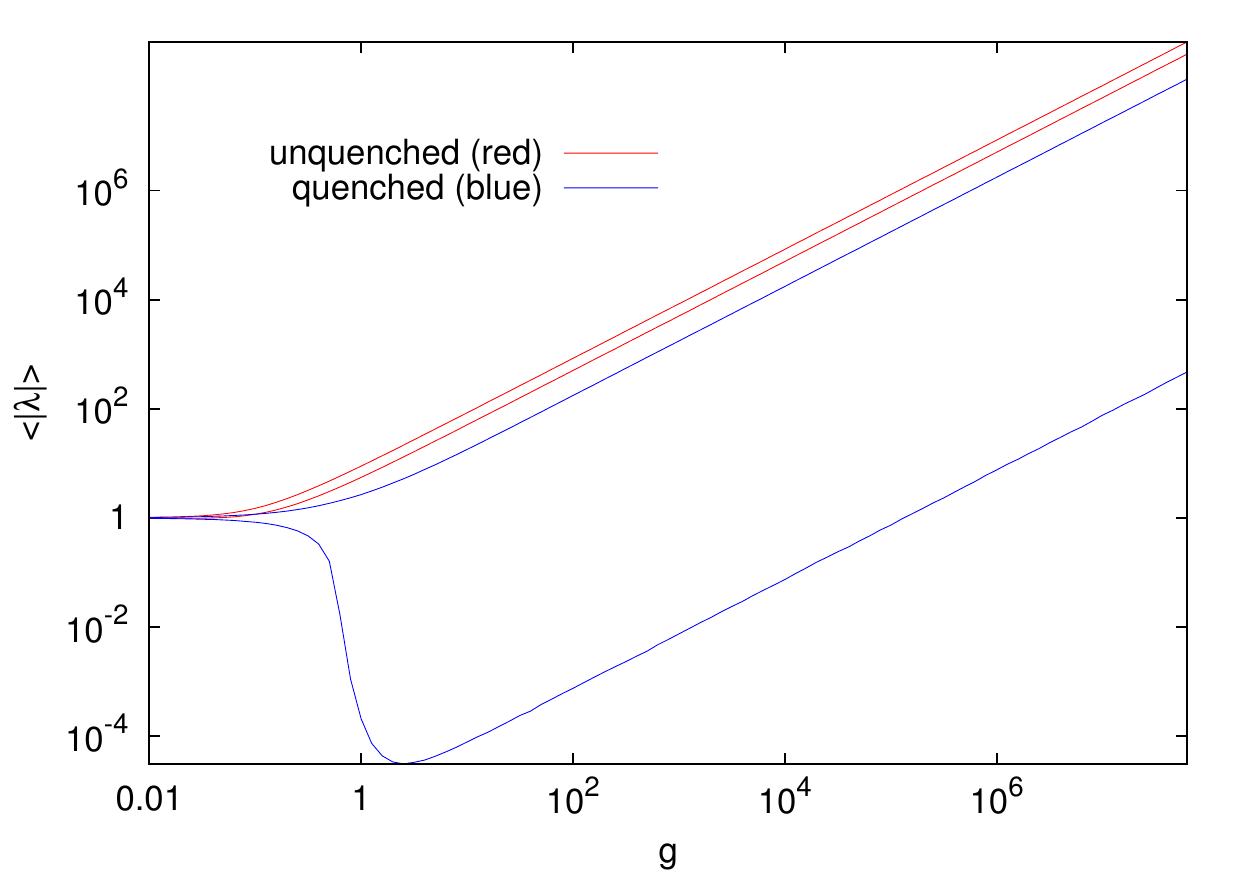}
\caption{$\lambda_{\rm min}$ and $\lambda_{\rm max}$ in both a quenched
and an unquenched simulation for $\kappa=0$ 
(see also Fig.~\ref{lambda_min_vs_g_various_L})
\label{lambda_min_unq}}
\end{figure}
\par\end{center}

\section{Summary and perspectives}

We have analyzed in the present paper the appearance of very small
eigenvalues of Dirac operator in a Yukawa theory with Wilson fermions.
The results obtained lead to the conclusion that at finite
volume and within the quenched approximation, these small eigenvalues
are present in an entire region of the phase space.
This indicates the existence of an ill conditioned region, not just
an ill conditioned line, for example the entire $\kappa=0$ line
is ill conditioned in the quenched approximation. Moreover
the size effects are exponentially large and consequently
a numerical calculation can give apparently correct results,
which would not survive the infinite volume limit.
In other words it does not seem possible to determine numerically
the ill conditioned region.
The origin of this difficulty could be simply the choice of the
discretization, or it could be the non validity of quenched approximation.
This hypothesis is supported by a an unquenched calculation for
$\kappa=0$, that is nowhere ill conditioned.
But it could also be a problem of the Yukawa model itself. Indeed
the Yukawa model is not a gauge model and there is
no protection against spurious low eigenvalues
like in QCD~\cite{protection}. 
%Finally we note that the region
%relevant for the physics is not the entire parmeter space, but
%only the vicinity of the point of the critical line where the ratio
%of the boson and fermion mass has the physical value.
%It is possible that this vicinity is not in the ill-conditionned region.

In this context we feel that the model should be studied without 
the quenched approximation. However a boson self coupling term
$\lambda \phi^4$ has to be added to the Lagrangian to ensure renormalizability.
%however the Lagrangian has to be modified to ensure renormalizability. 
This work is in progress.

%%%%%%%%%%%%%%%%%%%%%%%%%%%%%%%%%%%%%%%%%%%%%%%%%%%%%%%%%%%%
\appendix

\section{Average of extreme eigenvalues for $\kappa=0$}
\label{AppendixA}
In this appendix we show Eq.~\ref{a} and Eq.~\ref{aa}.
Since the $\phi_x$ are normally distributed with zero mean
and variance given by Eq.~\ref{eq:variance} the integrated probability
distribution of $|\lambda|$ is
$$
F_{|\Lambda|}(|\lambda|) = \frac{1}{2}\left( 
{\rm erf}(\frac{|\lambda|-1}{\sqrt2 g\sigma}) +
{\rm erf}(\frac{|\lambda|+1}{\sqrt2 g\sigma})\right)
$$
where $\sigma$ is the variance of the $\phi_x$.
Then from the definition of the min and
after an integration by part, one gets
\begin{eqnarray}
\left< | \lambda_{\rm min} |\right> &=& \int_0^\infty {(1-F_{|\Lambda|}(x))^N dx} 
\label{b} \\
\left< | \lambda_{\rm max} |\right> &=& \int_0^\infty {1-(F_{|\Lambda|}(x))^N dx} 
\label{c}
\end{eqnarray}
introducing $\phi(y)=1-F_{\sigma}(y)$ we have 
$$
\left\langle \lambda_{\mbox{min}}\right\rangle =\frac{1}{N}\int_{0}^{\infty}\left(\phi(\frac{y}{N})\right)^{N}dy
$$
Since
$$
\phi(h)=1-\sqrt{\frac{2}{\pi}}\frac{1}{\sigma}\exp-\frac{1}{2\sigma^{2}}h+O(h^{3})
$$
then
$$
\phi(\frac{x}{N})^{N}\rightarrow\exp\left(-x\sqrt{\frac{2}{\pi}}\frac{1}{g\sigma}e^{-\frac{1}{2g\sigma^{2}}}\right)
$$
so
$$
N\lambda_{\min}=\int_{0}^{\infty}\exp\left(-x\sqrt{\frac{2}{\pi}}\frac{1}{\sigma}e^{-\frac{1}{2\sigma^{2}}}\right)dx
$$
yielding Eq. \ref{eq:lambdamin}

We now study the behavior of $\lambda_{\mbox{max}}$. 
When $N$ is large the integrand in Eq.~\ref{c} tends 
to a step function equal to one for $x<x^{\star}$
and equal to zero for $x>x^{\star}$. One can
estimate $x^{\star}$ as the unique zero of the second derivative
of the integrand. Since $x^{\star}$ grows
when $N$ grows, one can replace $x-1$ and $x+1$ by $x$ in the
equation $\frac{d^{2}}{dx^{2}}F_{|\Lambda|}(x)^{N}=0$ yielding the equation
$$
N= \sqrt \frac{\pi}{2} 
\frac{\left\langle \lambda_{\rm max}\right\rangle}{g\sigma}
\exp \frac{1}{2} \left(\frac{\left\langle \lambda_{\rm max}\right\rangle}
{g\sigma}\right)^2
$$
from which Eq.~\ref{aa} follows.

\section{Estimators of boson correlator}
\label{appendixB}

\begin{flushleft}
The three estimators give the same correlator
\begin{equation}
C(t)=\frac{1}{L}\sum_{k_4}\frac{1}{\mu^{2}+\hat{k}_4^{2}} e^{i k_4 t}
\end{equation}
However the variances are different:
\begin{eqnarray*}
\sigma_{0}^{2}=\left\langle C_{0}^{2}(t)\right\rangle -\left\langle C_{0}(t)\right\rangle ^{2} & = & C^{2}(t)+A_{\mu}(L)C(t)\\
\sigma_{1}^{2}=\left\langle C_{1}^{2}(t)\right\rangle -\left\langle C_{1}(t)\right\rangle ^{2} & = & C^{2}(t)+C(0)C(t)\\
\sigma_{2}^{2}=\left\langle C_{2}^{2}(t)\right\rangle -\left\langle C_{2}(t)\right\rangle ^{2} & = & B_{\mu,L}(t)
\end{eqnarray*}
with
\par\end{flushleft}

\begin{flushleft}
\begin{eqnarray*}
A_{\mu}(L) & = & \sigma^2\ L^{3} \\
B_{\mu,L}(t) & = & \frac{1}{L^{2}}\sum_{k}\frac{1+e^{2\imath kt}}{\left(\mu^{2}+\hat{k_4^{2}}\right)^{2}}
\end{eqnarray*}
where $\sigma$ is defined in the text Eq.~\ref{eq:variance}.
Consequently one find $\sigma_{0}^{2}$ diverges as $L^{3}$, $\sigma_{1}^{2}$
tends to a finite value and $\sigma_2^{2}$ goes to zero as $\frac{\alpha_{\mu}}{\sqrt{L}}$
with $\alpha_{\mu}\sim\frac{1}{\mu^{4}}\quad\mu\rightarrow0$. Only
the third estimators $C_2(t)$ is self-averaging.
\par\end{flushleft}

\begin{acknowledgments}
We acknowledge J. Carbonell, M. Papinutto and O. Pene for many 
scientific discussions. 
This work was partially financed by Spanish projet FIS2010-18256.
\end{acknowledgments}

\end{document}